\documentclass[twocolumn,showpacs,floatfix, aps, nofootinbib]{revtex4}
\usepackage{amssymb,amsmath}
\usepackage[dvips]{graphics,color}
\usepackage{epsfig}\usepackage{float}
\newcommand{\ba}{\begin{eqnarray}}
\newcommand{\ea}{\end{eqnarray}}
\newcommand{\bege}{\begin{equation}}
\newcommand{\bpartial}{\mathop{\partial\kern -4pt\raisebox{.8pt}{$|$}}}
\newcommand{\enge}{\end{equation}}
\newcommand{\beq}{\begin{eqnarray}}
\newcommand{\benu}{\begin{enumerate}}

\newcommand{\enu}{\end{enumerate}}
\newcommand{\eeq}{\end{eqnarray}}
\newcommand{\pa}{\partial}

\newcommand{\RR}{\mathbb{R}}

\newcommand{\mk}{\mathfrak}

\newcommand{\me}{\frac{1}{2}}
\begin{document}

\title{Braneworld Remarks in Riemann-Cartan Manifolds}
\author{J. M. Hoff da Silva}
\email{hoff@ift.unesp.br} \affiliation{Instituto de F\'{\i}sica
Te\'orica, Universidade Estadual Paulista, Rua Pamplona 145
01405-900 S\~ao Paulo, SP, Brazil}
\author{R. da Rocha}
\email{roldao.rocha@ufabc.edu.br} \affiliation{ Centro de
Matem\'atica, Computa\c c\~ao e Cogni\c c\~ao, Universidade
Federal do ABC, 09210-170, Santo Andr\'e, SP, Brazil}


\begin{abstract}
We analyze the projected effective Einstein equation in a
$4$-dimensional arbitrary manifold embedded in a $5$-dimensional
Riemann-Cartan manifold. The Israel-Darmois matching conditions
are investigated, in the context where the torsion discontinuity
is orthogonal to the brane. Unexpectedly, the presence of torsion
terms in the connection does not modify such conditions
whatsoever, despite of the modification in the extrinsic curvature
and in the connection. Then, by imposing the
$\mathbb{Z}_{2}$-symmetry, the Einstein equation obtained via
Gauss-Codazzi formalism is extended, in order to now encompass the torsion terms.
We also show that the factors involving contorsion change
drastically the effective Einstein equation on the brane, as well
as the effective cosmological constant.
\end{abstract}
\maketitle

\section{Introduction}

In the last years there has been an increasing interest in large
extra dimension models \cite{GVALI}, mainly due to the
developments in string theory \cite{HW}, but also to the
possibility of the hierarchy problem explanation, presented for
instance in Randall-Sundrum and Ho$\check{\mathrm r}$ava-Witten
braneworld scenarios \cite{HW,RS}. In particular, the
Randall-Sundrum braneworld model \cite{RS} is effectively
implemented in a 5-dimensional manifold (where there is one warped
extra dimension) and it is based on a 5-dimensional reduction of
Ho$\check{\mathrm r}$ava-Witten theory \cite{nois,ROY}. In
Randall-Sundrum models, our universe is described by an infinitely
thin membrane
--- the brane. One attempt of explaining why gravity is so weak is
by trapping the braneworld in some higher-dimensional spacetime
--- the bulk --- wherein the brane is considered as a submanifold. For
instance, the observable universe proposed by Randall and Sundrum,
in one of their two models, can be described as being a brane
embedded in an AdS$_5$ bulk. There are several analogous models,
which consider our universe as a $D$-dimensional braneworld
embedded in a bulk of codimension one. In some models, there are
some modifications in the scenario that allow the presence of a
compact dimension on the brane \cite{PROP}. It gives rise to the
so called hybrid compactification.

As a crucial formal pre-requisite to try to describe gravity in a
braneworld context, the bulk is imposed to present codimension one
--- in relation to the brane. There is a great amount of results
applying the Gauss-Codazzi (GC) formalism \cite{WALD} in order to
derive the properties of such braneworld (see \cite{JP,reff} and
references therein). In the case where the bulk has two more
dimensions than the brane, the GC formalism is no longer useful,
since the concept of a thin membrane is meaningless, in the sense
that it is not possible to define junction conditions in
codimension greater than one. In such case the addition of a
Gauss-Bonnet term seems to break the braneworld apparent sterility
\cite{RUTH}. For higher codimensions, the situation is even worse.

Going back to the case of one non-compact extra dimension, after
expressing the Einstein tensor in terms os the stress tensor of
the bulk and extrinsic curvature corrections, it is necessary to
develop some mechanism to explore some physical quantities on the
brane. In order that the GC formalism to be useful, we must be
able to express the quantities in the limit of the extra dimension
going to zero --- at the point where the brane is located. Using
this procedure,  two junction requirements \cite{ISRAEL}, which
are the well known Israel-Darmois matching conditions, emerge.

A formalism where a manifold, endowed with a connection presenting
non-null torsion, is often required to describe physical theories
that are more general in many aspects. For instance, torsion is
essential when the description of fermionic matter coupled to
gravity is considered, and corrections of higher order in the
Einstein-Hilbert Lagrangian imply the presence of torsion in the
theory. Also, in contrast to the Yang-Mills formalism, in the
Poincar\'e gauge theory it is possible to construct an invariant
action which is linear in field derivatives. This gives rise to
the Einstein-Cartan theory as an immediate generalization of
General Relativity (GR) in a Riemann-Cartan manifold
\cite{kibble}. In the absence of matter fields, Einstein-Cartan
theory is equivalent to GR, and the invariant action in this case
is exactly the Hilbert-Palatini action. Moreover, torsion also
emerges in the interface between GR and gravity via string theory
at low energy. In this vein, it seems quite natural to explore
some aspects of braneworld models in the presence of torsion, which 
can be thought of as appearing in the theory as part of the connection gauge field
and work in the Palatini formalism in the representation of
orthonormal bases. In this case, the continuity of the projection of the connection field along the brane can 
be shown to follow from the consistency of the variational problem in
the presence of the brane. For instance, as regards the variational problem in the presence of the brane one can look into the paper \cite{kkk}.

This paper is organized as follows: after presenting some
geometric preliminaries involving Riemann-Cartan spacetimes in
Section II, we introduce the concept of torsion in the context of
GR and the Israel-Darmois matching conditions are investigated in
the presence of torsion, in an approach that is similar to the
formalism exhibited in reference \cite{PM}. In Section III,
junction conditions are investigated in the context where the
torsion discontinuity is {orthogonal} to the brane. Then, in
Section IV the Gauss-Codazzi formalism is used in order to
establish the role and implications of torsion terms in the
braneworld framework scenario.

\section{Braneworld preliminaries}

In this Section, we shall proceed as in ref. \cite{PM} presenting
the fudamental setup necessary to develop the formalism concerning
the matching conditions with torsion in the next Section, as well
as the application of GC formalism in the last
Section.\footnote{For a complete exposition concerning arbitrary
manifolds and fiber bundles, see, e.g, \cite{jjj, naka, koni,
moro, rowa}.}

Hereon $\Sigma$ denotes a $D$-dimensional Riemann-Cartan manifold
modelling a brane embedded in a bulk, denoted by $M$. A vector
space endowed with a constant signature  metric, isomorphic to
 $\RR^{D+1}$, can be identified at a point $x\in M$ as being the space
$T_xM$ tangent to $M$, where $M$ is locally diffeomorphic to its
local foliation $\RR\times\Sigma$. There always exists a 1-form
field $n$, normal to $\Sigma$, which can be locally interpreted
--- in the case where $n$ is timelike --- as being cotangent
to the worldline of observer families, i.e., the dual reference
frame relative velocity associated with such observers.

Denote $\{e_{a}\}$ ($a = 0,1,\ldots,D$) a basis for the tangent
space $T_x\Sigma$ at a point $x$ in $\Sigma$, and naturally the
cotangent space at $x\in\Sigma$ has an orthonormal basis $\{e^a\}$
such that $e^{a}(e_{b}) = \delta^{a}_{b}$. A reference frame at an
arbitrary point in the bulk is denoted by $\{e_{\alpha}\}$
  ($\alpha=0,1,2,\ldots,D+1$).  When a local coordinate chart is chosen, it is
possible to represent $e_{\alpha} = \partial/\partial x^{\alpha}
\equiv
\partial_{\alpha}$ and $e^{\alpha} = dx^{\alpha}$. The 1-form field orthogonal to the
sections of $T\Sigma$ --- the tangent bundle of $\Sigma$ --- can
now be written as $n = n_{\alpha} e^{\alpha}$,  and consider the
Gaussian coordinate $\ell$ orthogonal to the section of $T\Sigma$,
indicating how much an observer move out the $D$-dimensional brane
into the $(D+1)$-dimensional bulk.  A vector field $v = v^{\alpha}
e_{\alpha}$ in the bulk is split in components in the brane and
orthogonal to the brane, respectively as
 $v = v^{a} e_{a} + \ell e_{D+1}$. Since the bulk is endowed with a
non-degenerate bilinear symmetric form $g$ that can be written in
a coordinate basis as
 $g = g_{\alpha \beta}dx^{\alpha} \otimes dx^{\beta}$, the components of the metric
in the brane and on the bulk are  denoted respectively by
$q_{\alpha\beta}$ and
 $g_{\alpha\beta}$, and related by
\begin{equation}\label{neo}
g_{\alpha\beta} = q_{\alpha\beta} + n_{\alpha} n_{\beta}.
\end{equation}

The displacement away from the hypersurface, along one fixed but
arbitrary geodesic, is given by $d x^{\alpha} = n^{\alpha} d
\ell$, and in particular the expression $n_{\alpha} dx^{\alpha} =
d\ell$ implies that $n^{\alpha} n_{\alpha}=\pm 1$, where $+1$
corresponds to a spacelike braneworld $\Sigma$, and $-1$
corresponds to a timelike $\Sigma$. The 1-form field $n$
orthogonal to $\Sigma$,  in the direction of increasing $\ell$ is
given by $n =  (\pa_{\alpha} \ell)\,e^{\alpha}$, and its covariant
components are explicitly given by $n_\alpha = \pa_{\alpha} \ell$.
Without loss of generality a timelike hypersurface $\Sigma$ is
taken, where a congruence of geodesics goes across it.  Denoting
the proper distance (or proper time) along these geodesics by
$\ell$, it is always possible to put $\ell=0$ on $\Sigma$.

Denoting $\{x^\alpha\}$ a chart on both sides of the brane, define
another chart $\{y^a\}$ \emph{on} the brane. Here the same
notation used in \cite{PM} is adopted, where Latin indices are
used for hypersurface coordinates and Greek indices for
coordinates in the embedding spacetime. The brane can be
parametrized by $x^\alpha=x^\alpha(y^a)$, where the parametric
index $a$ runs over the dimensions of the hypersurface --- not
being a spacetime index --- and the vierbein
$h^\alpha_a:=\frac{\pa x^\alpha}{\pa y^a} $ satisfy $h^\alpha_a
n_\alpha=0$. For displacements on the brane, it follows that \beq
g &=& g_{\alpha\beta}\,d x^\alpha \otimes d x^\beta =
g_{\alpha\beta}\,\Big(\frac{\pa x^\alpha}{\pa y^a}\,d y^a
\Big)\otimes
\Big(\frac{\pa x^\beta}{\pa y^b}\,d y^b \Big) \nonumber\\
&=& q_{ab}\,d y^a \otimes d y^b, \eeq and so the induced metric
components $q_{ab}$ on $\Sigma$ is related to $g_{\alpha\beta}$ by
$q_{ab} = g_{\alpha\beta}\,h^\alpha_a h^\beta_b.$

Denoting by $ [A] = {}\lim_{\ell\rightarrow 0^+}(A) -
\lim_{\ell\rightarrow 0^-}(A) $ the change in a differential form
field $A$ across the braneworld $\Sigma$ (wherein $\ell = 0$), the
continuity of the chart $x^\alpha$ and $\ell$ across $\Sigma$
implies that $n_\alpha$ and $h^\alpha_a$ are continuous, or,
equivalently,  $[n_\alpha]=[h^\alpha_a]=0$.

Now, using the Heaviside distribution $\Theta(\ell)$
properties\footnote{$\delta(\ell)$ is the Dirac distribution.}
\[
\Theta^2(\ell)=\Theta(\ell), \qquad \Theta(\ell)\Theta(-\ell)=0,
\qquad \frac{d}{d \ell}\,\Theta(\ell) = \delta(\ell),
\]
the metric components $g_{\alpha\beta}$ can be written as
distribution-valued tensor components
\[
g_{\alpha\beta}=\Theta(\ell)\,g^+_{\alpha\beta}+\Theta(-\ell)\,g^-_{\alpha\beta},
\]
where $g^+_{\alpha\beta}$ ($g^-_{\alpha\beta}$) denotes the metric
on the $\ell>0$ ($\ell<0$) side of $\Sigma$. Differentiating the
above expression, it reads
\[
\pa_\gamma g_{\alpha\beta}=\Theta(\ell)\,\pa_\gamma
g^+_{\alpha\beta}+\Theta(-\ell)\,\pa_\gamma g^-_{\alpha\beta} +
\delta(\ell)[g_{\alpha\beta}]n_\gamma.
\]
The last term is singular; moreover, this term creates problems
when we compute the Christoffel symbols by generating the product
$\Theta(\ell)\delta(\ell)$, which is not defined as a
distribution. It can be shown that the condition
$[g_{\alpha\beta}]=0$ must be imposed for the connection to be
defined as a distribution\footnote{Basically, if the condition
 $[g_{\alpha\beta}]=0$ is not imposed, there appears the product
$\Theta\delta$, which is not well defined in the Levi-Civita part
of the connection. }, also implying  the `first' junction
condition $[h_{ab}]$.

Besides a curvature associated with the connection that endows the
bulk, in a Riemann-Cartan manifold the torsion associated with the
connection is in general non zero. Its components can be written
in terms of the connection components
$\Gamma^{\rho}{}_{\beta\alpha}$ as
\begin{equation}
T^{\rho}{}_{\alpha\beta} = \Gamma^{\rho}{}_{\beta\alpha} -
\Gamma^{\rho}{}_{\alpha\beta}. \label{tor}
\end{equation}
The general connection components are related to the Levi-Civita
connection components
${\stackrel{\circ}{\Gamma}}{}^{\rho}{}_{\alpha\beta}$ ---
associated with the spacetime metric $g_{\alpha\beta}$ components
--- through $\Gamma^{\rho}{}_{\alpha\beta} =
{\stackrel{\circ}{\Gamma}}{}^{\rho}{}_{\alpha\beta} +
K^{\rho}{}_{\alpha\beta}$, where $K^{\rho}{}_{\alpha \beta} =
\textstyle{\frac{1}{2}} \left( T_{\alpha}{}^{\rho}{}_{\beta} +
T_{\beta}{}^{\rho}{}_{\alpha} - T^{\rho}{}_{\alpha \beta} \right)$
denotes the contortion tensor components. It must be emphasized
that curvature and torsion are properties of a connection, not of
spacetime. For instance, the Christoffel and the general
connections present different curvature and torsion, although they
endow the very same manifold.

Now the distribution-valued Riemann tensor is calculated, in order
to find the `second' junction condition  --- the Israel matching
condition. From the Christoffel symbols, it reads $
\Gamma^\alpha_{\;\,\beta\gamma} = \Theta(\ell)
\Gamma^{+\alpha}_{\;\;\,\beta\gamma}+\Theta(-\ell)\Gamma^{-\alpha}_{\;\;\,\beta\gamma},
$ where $\Gamma^{\pm\alpha}_{\;\;\,\beta\gamma}$ are the
Christoffel symbols obtained from $g^\pm_{\alpha\beta}$.  Thus
\[
\pa_\delta\Gamma^\alpha_{\;\,\beta\gamma} =
\Theta(\ell)\pa_\delta\Gamma^{+\alpha}_{\;\;\;\,\beta\gamma}
+\Theta(-\ell)\pa_\delta\Gamma^{-\alpha}_{\;\;\;\,\beta\gamma} +
\delta(\ell)[\Gamma^\alpha_{\;\,\beta\gamma}]n_\delta,
\]
and the Riemann tensor is given by $
R^\alpha_{\beta\gamma\delta}=\Theta(\ell)R^{+\alpha}_{\;\;\;\,\beta\gamma\delta}+\Theta(-\ell)R^{-\alpha}_{\;\;\;\,\beta\gamma\delta}
+\delta(\ell)A^\alpha_{\;\,\beta\gamma\delta}, $ where $
A^\alpha_{\;\,\beta\gamma\delta}=[\Gamma^\alpha_{\;\,\beta\delta}]n_\gamma-[\Gamma^\alpha_{\;\,\beta\gamma}]n_\delta$.

The next step is to find an explicit expression for the tensor
$A^\alpha_{\;\,\beta\gamma\delta}$.

\section{Junction conditions with the torsion discontinuity orthogonal to the brane}

Observe that the continuity of the metric across $\Sigma$ implies
that the tangential derivatives of the metric must be also
continuous. If $\partial_\gamma g_{\alpha\beta} \equiv
g_{\alpha\beta, \gamma}$ is indeed discontinuous, this
discontinuity must be directed along the normal vector $n^\alpha$.
It is therefore possible to write
\[
[g_{\alpha\beta,\gamma}]=\kappa_{\alpha\beta}n_\gamma,
\]
for some tensor $\kappa_{\alpha\beta}$ (given explicitly by
$\kappa_{\alpha\beta}=[g_{\alpha\beta,\gamma}]n^\gamma$). Then it
follows that
\[
[{\stackrel{\circ}{\Gamma}}{}^\alpha_{\;\,\beta\gamma}] =
\me\,(\kappa^\alpha_{\;\,\beta}
n_\gamma+\kappa^\alpha_{\;\,\gamma} n_\beta - \kappa_{\beta\gamma}
n^\alpha),
\]
and supposing that the discontinuity in the torsion terms obey the
same rule as the discontinuity of $[g_{\alpha\beta,\gamma}]$, i.
e., that
$[T^{\alpha}_{\;\,\beta\gamma}]=\zeta^{\;\alpha}_{\beta}n_{\gamma}$,
it reads \beq [K^\alpha_{\;\,\beta\gamma}] &=&
\me\,(\zeta^{\;\,\alpha}_\beta n_\gamma+\zeta^{\;\,\alpha}_\gamma
n_\beta - \zeta^{\alpha}_{\;\,\beta} n_\gamma).\label{ca} \eeq The
components $\kappa_{\rho\sigma}$ emulate an intrinsic property of
the brane itself. The torsion is continuous along the brane, and
if there is some discontinuity, it is proportional to the extra
dimension. Such proportionality is given, in principle, by another
quantity  $\zeta^{\;\alpha}_{\beta}$ related to the brane. After
these considerations, it follows that \beq
[\Gamma^\alpha_{\;\,\beta\gamma}] &=&
 \me\,((\kappa^\alpha_{\;\,\beta} + \zeta^{\;\,\alpha}_\beta -
\zeta^{\alpha}_{\;\,\beta}) n_\gamma
 +(\kappa^\alpha_{\;\,\gamma}+\zeta^{\;\,\alpha}_\gamma) n_\beta - \kappa_{\beta\gamma}
 n^\alpha),\nonumber
\eeq and hence \ba
A^\alpha_{\;\,\beta\gamma\delta}&=&\left.\frac{1}{2}\,(\kappa^\alpha_{\;\,\delta}
n_\beta n_\gamma - \kappa^\alpha_{\;\,\gamma} n_\beta n_\delta -
\kappa_{\beta\delta} n^\alpha n_\gamma
+\kappa_{\beta\gamma}n^\alpha n_\delta)\right. \nonumber\\&&+
\left. \me(\zeta_\delta^{\;\,\alpha}n_\beta n_\gamma -
\zeta_\gamma^{\;\,\alpha}n_\beta n_\delta).\right. \ea Denoting
$\kappa = \kappa^\alpha_{\;\,\alpha}$ and $\zeta =
\zeta^\beta_{\;\,\beta}$, and suitably contracting two indices, it
reads \ba
A_{\beta\delta}&=&\left.\frac{1}{2}(\kappa^\alpha_{\;\,\delta}
n_\beta n_\alpha -\kappa n_\beta n_\delta - \kappa_{\beta\delta} +
\kappa_{\beta\alpha} n^\alpha n_\delta)\right.\nonumber
\\&+&\left. \me(\zeta_\delta^{\;\,\alpha}n_\beta n_\alpha - \zeta
n_\beta n_\delta),\right. \eeq and also \beq A =
g^{\beta\delta}A_{\beta\delta} = (\kappa_{\alpha\delta} n^\alpha
n^\delta - \kappa) + \me(\zeta_{\delta\alpha}n^\delta n^\alpha -
\zeta). \nonumber\eeq

The $\delta$-function part of the Einstein tensor
$G_{\alpha\beta}:=R_{\alpha\beta}-\me g_{\alpha\beta}R$ is given
by \beq \label{Seq}
S_{\beta\delta} &=& A_{\beta\delta}-\me g_{\beta\delta}A\nonumber\\
&=& \frac{1}{2}\,(\kappa_{\;\,\delta}^{\alpha}n_\beta n_\alpha -
\kappa n_\beta n_\delta  -
\kappa_{\beta\delta} + \kappa_{\beta\alpha}n^\alpha n_\delta \nonumber\\
&& - g_{\beta\delta}(\kappa_{\rho\sigma}n^\rho n^\sigma -
\kappa))+ \frac{1}{2}\,(\zeta_{\delta}^{\;\,\alpha}n_\beta
n_\alpha - \zeta
n_\beta n_\delta) \nonumber\\
&& -\frac{1}{4} g_{\beta\delta}(\zeta_{\rho\sigma}n^\rho n^\sigma
- \zeta)).\label{opa} \eeq On the other hand, the total
stress-energy tensor has the form
\[
\pi^\mathrm{\,total}_{\alpha\beta}=\Theta(\ell)\pi^+_{\alpha\beta}+\Theta(-\ell)\pi^-_{\alpha\beta}+\delta(\ell)\pi_{\alpha\beta},
\]
where $\pi^+_{\alpha\beta}$ and $\pi^-_{\alpha\beta}$ represent
the bulk stress-energy in the regions where $\ell> 0$ and $\ell <
0$ respectively, while $\pi_{\alpha\beta}$ denotes the
stress-energy localized on the hypersurface $\Sigma$ itself. From
the Einstein equations, it follows that
$\pi_{\alpha\beta}=(G_{N})^{-1} S_{\alpha\beta}$.

Note that, since $\pi_{\alpha\beta}$ is tangent to the brane, it
follows that $\pi_{\alpha\beta}n^{\beta}=0$. However, from
Eq.(\ref{Seq}) the following equation \beq 4 G_{N}
\pi_{\alpha\beta}n^\beta&=&\frac{1}{2}(\zeta_{\rho\sigma}n^\rho
n^\sigma - \zeta)n_\alpha \nonumber\\
&=& -\frac{1}{2}\mk \zeta_{\rho\sigma}{q}^{\rho\sigma} n_\alpha,
\eeq is derived, which means that, in order to keep the
consistence of the formalism, one has to impose
$\zeta_{\rho\sigma}q^{\rho\sigma}=0$, and the last term of
Eq.(\ref{opa}) vanishes. Note that  $\pi_{\alpha\beta}$ can be
expressed by $\pi_{ab}=\pi_{\alpha\beta}h^\alpha_a h^{\beta}_b$,
just using the $h_a^\alpha$ vierbein introduced in the previous
Section. So, taking into account that
$\pi_{\alpha\beta}=(G_{N})^{-1} S_{\alpha\beta}$ and
Eq.(\ref{opa}), it reads \cite{PM} \beq 4 G_{N}
\pi_{ab}=-\kappa_{\alpha\beta} h^\alpha_a
h^\beta_b+q^{rs}\kappa_{\mu\nu}h^\mu_r h^\nu_s q_{ab}.\eeq
Finally, relating the jump in the extrinsic curvature to
$\kappa_{\rho\sigma}$, via the covariant derivative associated to
$q_{\alpha\beta}$, the following expression can be obtained  from
Eq.(\ref{ca}):  \ba
[\nabla_{\alpha}n_\beta]&=&\left.\frac{1}{2}(\kappa_{\alpha\beta}-\kappa_{\gamma\alpha}n_{\beta}n^{\gamma}-\kappa_{\gamma\beta}n_\alpha
n^\gamma)\right.\nonumber\\&+&\left.\frac{1}{2}(\zeta^{\;\,\gamma}_\alpha
n_\beta+\zeta^{\;\,\gamma}_\beta n_\alpha -
\zeta^{\gamma}_{\;\,\alpha} n_\beta)n_\gamma .\right.\ea However,
it is clear that this jump of the extrinsic curvature across the
brane, $[\nabla_{\alpha}n_\beta]\equiv [\Xi_{\alpha\beta}]$, can
be also decomposed in terms of $h_a^\alpha$ vectors, leading to
\beq [\Xi_{ab}]=\frac{1}{2}\kappa_{\alpha\beta}h^\alpha_a
h^\beta_b .\eeq Hence, after all, it follows that \beq 2G_{N}
\pi_{ab}=-[\Xi_{ab}]+[\Xi]q_{ab}.\label{vai}\eeq It means that the
second matching condition is absolutely the same that is valid
without any torsion term. So, there is no difference in both
junctions conditions within the context of a Riemann-Cartan
manifold, which is an unexpected characteristic, since the torsion
terms are directly related to the extrinsic curvature
$(\nabla_{\alpha}n_{\beta})$ and effectively modify the
connection.

\section{The projected equations on the brane}

We have investigated the matching conditions in the presence of
torsion terms, and under the assumptions of discontinuity across
the brane. Surprisingly, both junctions conditions are shown to be
the same as the usual case
($\Gamma^{\rho}_{\;\,\alpha\beta}={\stackrel{\circ}{\Gamma}}{}^{\rho}_{\;\,\alpha\beta}$).
We remark that, since the covariant derivative changes by torsion,
the extrinsic curvature is also modified, and then the
conventional arguments point in the direction of some modification
in the matching conditions. However, it seems that the r\^ole of
torsion terms in the braneworld picture is restricted to the
geometric part of effective Einstein equation on the brane. More
explicitly, looking at the equation that relates the Einstein
equation in four dimensions with bulk quantities (see, for example
\cite{JP}) we have
\begin{eqnarray}
^{(4)}\!G_{\rho\sigma}&=&\left.\frac{2k_{5}^{2}}{3}\Bigg(T_{\alpha\beta}q_{\rho}^{\;\alpha}q_{\sigma}^{\;\beta}+(T_{\alpha\beta}
n^{\alpha}n^{\beta}-\frac{1}{4}T)q_{\rho\sigma}\Bigg)\right.\nonumber
\\&&+\left. \Xi\Xi_{\rho\sigma}-\Xi_{\rho}^{\;\alpha}\Xi_{\alpha\sigma}-
\frac{1}{2}q_{\rho\sigma}(\Xi^{2}-\Xi^{\alpha\beta}\Xi_{\alpha\beta})\right.
\nonumber
\\&&-\left.\;^{(5)}\!C^{\alpha}_{\;\beta\gamma\epsilon}n_{\alpha}n^{\gamma}q_{\rho}^{\;\beta}q_{\sigma}^{\;\epsilon},\right.\label{gde}
\end{eqnarray}
where $T_{\rho\sigma}$ denotes the energy-momentum tensor,
$\Xi_{\rho\sigma}=q_{\rho}^{\;\alpha}q_{\sigma}^{\;\beta}\nabla_{\alpha}n_{\beta}$
is the extrinsic curvature, $k_5$ denotes the 5-dimensional
gravitational constant, and
$^{(5)}\!C^{\alpha}_{\;\beta\rho\sigma}$ denotes the Weyl tensor.
By restricting to quantities evaluated on the brane, or tending to
the brane, we see that the only way to get some contribution from
torsion terms is via the term $^{(4)}\!G_{\rho\sigma}$, and also
via the Weyl tensor. In the light of  Section III it does not
intervene in the extrinsic curvature tending to the brane.
Actually, this fact makes the calculations easier when one tries
to apply it to a particular model, specially possessing
$\mathbb{Z}_{2}$-symmetry, to extract more information about the
r\^ole of the torsion in gravitational systems considered in
braneworld scenarios.

In order to explicit the influence of (con)torsion terms in the
projected equations on the brane, we shall to complete the GC
program, from five to four dimensions, to the case with torsion.
Note the by imposing the $\mathbb{Z}_{2}$-symmetry, the extrinsic
curvature reads
\ba\Xi^{+}_{\alpha\beta}=-\Xi^{-}_{\alpha\beta}=-2G_{N}\Big(\pi_{\alpha\beta}-\frac{q_{\alpha\beta}\pi^{\gamma}_{\gamma}}{4}\Big),\label{c1}\ea
in such way that Eq.(\ref{vai}) reads\footnote{Hereon, we remove
the $+$ and $-$ labels.}
\ba\Xi_{\alpha\beta}=-G_{N}\Big(\pi_{\alpha\beta}-\frac{q_{\alpha\beta}\pi^{\gamma}_{\gamma}}{4}\Big).\label{c2}\ea

Decomposing the stress-tensor associated with the
bulk\footnote{Note that the delta factor appearing in
$T_{\alpha\beta}=-\Lambda g_{\alpha\beta}+\delta S_{\alpha\beta}$
is necessary here, in order to localize the brane. In fact, this
type of decomposition is compatible with the Israel-Darmois
junction conditions. We remark that such a delta term can lead to
problems in a more complete cosmological scenario, but for the
purpose of this work there is not problem.} in
$T_{\alpha\beta}=-\Lambda g_{\alpha\beta}+\delta S_{\alpha\beta}$
and $S_{\alpha\beta}=-\lambda q_{\alpha\beta}+\pi_{\alpha\beta}$,
where $\Lambda$ is the bulk cosmological constant and $\lambda$
the brane tension, and substituting into Eq.(\ref{gde}) it follows
after some algebra\footnote{See, please, reference \cite{JP} for
all the details.}, \ba
^{(4)}\!G_{\mu\nu}=-\Lambda_{4}q_{\mu\nu}+8\pi
G_{N}\pi_{\mu\nu}+k_{5}^{4}Y_{\mu\nu}-E_{\mu\nu} ,\label{c3}\ea
where
$E_{\mu\nu}=^{(5)}\!C^{\alpha}_{\beta\gamma\sigma}n_{\alpha}n^{\gamma}q_{\mu}^{\beta}q_{\nu}^{\sigma}$
encodes the Weyl tensor contribution, $G_{N}=\frac{\lambda
k_{5}^{4}}{48\pi}$ is the analogous of the Newton gravitational
constant, the tensor $Y_{\mu\nu}$ is quadratic in the brane
stress-tensor and given by
$Y_{\mu\nu}=-\frac{1}{4}\pi_{\mu\alpha}\pi_{\nu}^{\alpha}+\frac{1}{12}\pi^{\gamma}_{\gamma}\pi_{\mu\nu}+\frac{1}{8}
q_{\mu\nu}\pi_{\alpha\beta}\pi^{\alpha\beta}-\frac{1}{2}q_{\mu\nu}(\pi^{\gamma}_{\gamma})^{2}$
and
$\Lambda_{4}=\frac{k_{5}^{2}}{2}\Big(\Lambda+\frac{1}{6}k_{5}^{2}\lambda^{2}\Big)$
is the effective brane cosmological constant.

Now, the contributions arising from the (con)torsion terms are
explicited in details. It is well known that the Riemann and Ricci
tensors, and the curvature scalar written in terms of torsion are
related with their partners, constructed with the usual metric
compatible Levi-Civita connection by \cite{SHA}
\begin{eqnarray}
R^{\lambda}_{\;\;\tau\alpha\beta}&=&\left.\mathring{R}^{\lambda}_{\;\;\tau\alpha\beta}+\nabla_{\alpha}K^{\lambda}_{\;\;\tau\beta}
-\nabla_{\beta}K^{\lambda}_{\;\;\tau\alpha}\right.\nonumber\\&&+\left.K^{\lambda}_{\;\;\gamma\alpha}K^{\gamma}_{\;\;\tau\beta}
-K^{\lambda}_{\;\;\gamma\beta}K^{\gamma}_{\;\;\tau\alpha},\right.
\label{c4}
\end{eqnarray}
\begin{eqnarray}
R_{\tau\beta}&=&\left.\mathring{R}_{\tau\beta}+\nabla_{\lambda}K^{\lambda}_{\;\;\tau\beta}-\nabla_{\beta}K^{\lambda}_{\;\;\tau\lambda}
\right.\nonumber\\&+&K^{\lambda}_{\;\;\gamma\lambda}K^{\gamma}_{\;\;\tau\beta}
-K^{\lambda}_{\;\;\tau\gamma}K^{\gamma}_{\;\;\lambda\beta}
\label{c5}
\end{eqnarray}
and
\begin{eqnarray}
R=\mathring{R}+2\nabla^{\lambda}K^{\tau}_{\;\;\lambda\tau}-K_{\tau\lambda}^{\;\;\;\lambda}K^{\tau
\lambda}_{\;\;\;\lambda}+K_{\tau\gamma\lambda}K^{\tau\lambda\gamma},\label{c6}
\end{eqnarray}
where the quantities $\mathring{X}$ are constructed with the usual
metric compatible Levi-Civita connection, and $\nabla$ denotes the
covariant derivative {\it without} torsion. Clearly such relations
holds in any dimension. Therefore, by denoting $D_{\mu}$ the
covariant 4-dimensional derivative acting on the brane, it is easy
to see that, from Eqs.(\ref{c4}),(\ref{c5}), and (\ref{c6}), the
Einstein tensor on the brane is given by
\begin{eqnarray}
^{(4)}\!G_{\mu\nu}&=&\left.
^{(4)}\!\!\mathring{G}_{\mu\nu}+D_{\lambda}\;^{(4)}\!K^{\lambda}_{\;\;\mu\nu}-D_{\nu}\;^{(4)}\!K^{\lambda}_{\;\;\mu\lambda}\right.\nonumber
\\&&+^{(4)}\!K^{\lambda}_{\;\;\gamma\lambda}\;^{(4)}K^{\gamma}_{\;\;\mu\nu}
-\;^{(4)}K^{\lambda}_{\;\;\mu\gamma}\;^{(4)}K^{\gamma}_{\;\;\lambda\nu}\nonumber\\&&\left.-q_{\mu\nu}\Big(
D^{\lambda}\;^{(4)}\!K^{\tau}_{\;\;\lambda\tau}+\frac{1}{2}\;^{(4)}\!K^{\;\;\;\lambda}_{\tau\lambda}
\;^{(4)}\!K^{\tau\gamma}_{\;\;\;\gamma}\right.\nonumber\\&&\left.+\frac{1}{2}\;^{(4)}\!K_{\tau\gamma\lambda}\;^{(4)}K^{\tau\gamma\lambda}\Big)\right..
\label{c7}
\end{eqnarray}
Note the appearance of terms multiplying the brane metric. As it
shall be seen, these terms compose a new effective cosmological
constant.

The $E_{\mu\nu}$ tensor can be expressed in terms of the bulk
contorsion terms $(K^{\mu}_{\;\;\nu\alpha})$ by
\begin{eqnarray}
\hspace{-.5cm}E_{\kappa\delta}&=&\left.\stackrel{\circ}{E}_{\kappa\delta}+\Big(\nabla_{\nu}K^{\mu}_{\;\;\alpha\beta}-\nabla_{\beta}K^{\mu}_{\;\;\alpha\nu}+
K^{\mu}_{\;\;\gamma\nu}K^{\gamma}_{\;\;\alpha\beta}\right.\nonumber\\&&-\left.K^{\mu}_{\;\;\gamma\beta}K^{\gamma}_{\;\;\alpha\nu}\Big)n_{\mu}n^{\nu}
q_{\kappa}^{\alpha}q_{\delta}^{\beta}-\frac{2}{3}(q_{\kappa}^{\alpha}q_{\delta}^{\beta}+n^{\alpha}n^{\beta}q_{\kappa\delta})\right.\nonumber\\&&\times
\left.
\Big(\nabla_{\lambda}K^{\lambda}_{\;\;\beta\alpha}-\nabla_{\alpha}K^{\lambda}_{\;\;\beta\lambda}+K^{\lambda}_{\;\;\gamma\lambda}K^{\gamma}_{\;\;\beta\alpha}
-K^{\sigma}_{\;\;\beta\gamma}K^{\gamma}_{\;\;\sigma\alpha}\Big)\right.\nonumber\\&&+\frac{1}{6}q_{\kappa\delta}\left.
\Big(2\nabla^{\lambda}K^{\tau}_{\;\;\lambda\tau}-K_{\tau\lambda}^{\;\;\;\lambda}K^{\tau\gamma}_{\;\;\;\gamma}+K_{\tau\gamma\lambda}
K^{\tau\lambda\gamma} \Big) \right. \label{c8}
\end{eqnarray} where $\nabla_{\mu}$ is the bulk covariant
derivative. Now, the explicit influence of the contorsion terms in
the Einstein brane equation can be appreciated. From
Eqs.(\ref{c3}), (\ref{c7}), and (\ref{c8}), it reads
\begin{widetext}
\begin{eqnarray}&&
\left.
\hspace{-1cm}^{(4)}\!\mathring{G}_{\mu\nu}+D_{\lambda}\;^{(4)}\!K^{\lambda}_{\;\;\mu\nu}-D_{\nu}\;^{(4)}\!K^{\lambda}_{\;\;\mu\lambda}+
^{(4)}\!K^{\delta}_{\;\;\gamma\delta}\;^{(4)}\!K^{\lambda}_{\;\;\mu\gamma}\;^{(4)}\!K^{\gamma}_{\;\;\lambda\nu}
=-\tilde{\Lambda}_{4}q_{\mu\nu}+8\pi
G_{N}\pi_{\mu\nu}+k_{5}^{4}Y_{\mu\nu}-\mathring{E}_{\mu\nu}+q_{\mu}^{\alpha}q_{\nu}^{\beta}\right.\nonumber
\\&&\hspace{-1cm}\times \left.\Bigg[\frac{2}{3}\Big(\nabla_{\lambda}
K^{\lambda}_{\;\;\beta\alpha}-\nabla_{\alpha}K^{\lambda}_{\;\;\beta\lambda}+K^{\sigma}_{\;\;\gamma\sigma}K^{\gamma}_{\;\;\beta\alpha}
-K^{\lambda}_{\;\;\beta\gamma}K^{\gamma}_{\;\;\lambda\alpha}\Big)-n_{\rho}n^{\sigma}\Big(\nabla_{\sigma}K^{\rho}_{\;\;\alpha\beta}-
\nabla_{\beta}K^{\rho}_{\alpha\sigma}+K^{\rho}_{\;\;\gamma\sigma}K^{\gamma}_{\;\;\alpha\beta}-K^{\rho}_{\;\;\gamma\beta}K^{\gamma}_{\;\;\alpha\sigma}
\Big)\Bigg] \right. \label{c9}
\end{eqnarray}
where the new effective cosmological constant is given by
\begin{eqnarray}
\tilde{\Lambda}_{4}&\equiv&\left.
\Lambda_{4}-D^{\lambda}\;^{(4)}\!K^{\tau}_{\;\;\lambda\tau}+\frac{1}{2}\;^{(4)}K_{\tau\alpha}^{\;\;\;\alpha}\;^{(4)}K^{\tau\lambda}_{\;\;\;\lambda}
-\frac{1}{2}\;\;^{(4)}K_{\tau\gamma\lambda}\;^{(4)}K^{\tau\lambda\gamma}-\frac{2}{3}n^{\alpha}n^{\beta}\Big(\nabla_{\lambda}K^{\lambda}_{\;\;\beta\alpha}-
\nabla_{\alpha}K^{\lambda}_{\;\;\beta\lambda}\right.\nonumber\\&&+\left.K^{\lambda}_{\;\;\gamma\lambda}K^{\gamma}_{\;\;\beta\alpha}
-K^{\sigma}_{\;\;\beta\gamma}K^{\gamma}_{\;\;\sigma\alpha}\Big)+\frac{1}{6}\Big(2\nabla^{\lambda}K^{\tau}_{\;\;\lambda\tau}-
K_{\tau\alpha}^{\;\;\;\alpha}K^{\tau\lambda}_{\;\;\;\lambda}+K_{\tau\gamma\lambda}K^{\tau\lambda\gamma}\Big).\label{c10}
\right.
\end{eqnarray}
\end{widetext}
Eqs. (\ref{c9}) and (\ref{c10}) enclose the main result of this
paper. From Eq.(\ref{c9}) it follows that the factors involving
both contorsion, in four and in five dimensions, change
drastically the effective Einstein equation on the brane, as well
as the effective cosmological constant. We shall comment this
important and remarkable result in the next Section.

\section{Concluding remarks and outlook}

There are some alternative derivations of the junction conditions
for a brane in a 5-dimensional bulk, when Gauss-Bonnet equations
are used to describe gravity \cite{deruele}. Also, Israel junction
conditions can be generalized for a wider class of theories by
direct integration of the field equations, where a specific
non-minimal coupling of matter to gravity suggests promising
classes of braneworld scenarios \cite{japa}.  In addition, it is
also possible to generalize matching conditions for cosmological
perturbations in a teleparallel Friedmann universe, following the
same lines as \cite{deru}.

In the case studied here, however, the matching conditions are not
modified by the inclusion of torsion terms in the connection. As
noted, it is a remarkable and unexpected characteristic. Besides,
all the development concerning the formalism presented is
accomplished in the context of braneworld models. In such
framework, the appearance of torsion terms is quite justifiable.
However, the fact that the matching conditions remain unalterable
in the presence of torsion is still valid in usual 4-dimensional
theories.

Once investigated the junction conditions, we have obtained, via
Gauss-Codazzi formalism, the Einstein effective projected equation
on the brane. If, on one hand, the torsion terms do not intervenes
in the usual Israel-Darmois conditions, on the other hand it
modifies drastically the brane Einstein equations. Eq.(\ref{c10})
shows up the strong dependence of the new effective cosmological
constant on the four and five-dimensional contorsion terms. It
reveals promising possibilities. For instance, by a suitable
behavior of such new terms, $\tilde{\Lambda}_{4}$ can be very
small. In a more complete scenario, $\tilde{\Lambda}_{4}$ could be
not even a constant. It must be stressed that these types of
modification in the projected Einstein equation also appear in
other models in modified gravity \cite{NDN}.

This paper intends to give the necessary step in order to formalize
the mathematical implementation of torsion terms in braneworld
scenarios. The application of our results are beyond the scope of
this work. We finalize, however, pointing out some interesting research
lines coming from the use of the results --- obtained in this paper --- in cosmological
problems.

The final result is very important from the cosmological viewpoint. It is clear that deviations of the usual
braneworld cosmology can be obtained from the analysis of phenomenological systems
in the light of Eq.(\ref{c9}). Physical aspects, more specifically
the analysis of cosmological signatures as found in ref.
\cite{OC}, arising from the combination of the extra dimensions and
torsion should be systematically investigated and compared with
usual braneworld models. The ubiquitous presence of torsion terms
leads, by all means, to subtle but important deviations of usual
braneworlds in General Relativity. For instance, the equation
(\ref{c9}) can be used as a starting point to describe the flat
behavior of galaxy rotational curves without claim for dark
matter. This last problem was already analyzed in the context of
brane worlds \cite{ROTGAL}, however the outcome arising from the
torsion terms has never been investigated. A systematic
comparative study between usual braneworld models and those braneworld models
embedded in an Einstein-Cartan manifold is, potentially,
interesting since it can lead us to new branches inside brane
physics. We shall address to those questions in the future.

\section*{Acknowledgment}

The authors thank to Prof. R. A. Mosna and the CQG Referee for fruitful suggestions.
J. M. Hoff da Silva thanks to CAPES-Brazil for financial support.

\end{document}